\documentclass{article}

\PassOptionsToPackage{numbers, compress}{natbib}

\usepackage[preprint]{neurips_2024}

\usepackage[utf8]{inputenc} 
\usepackage[T1]{fontenc}    
\usepackage{hyperref}       
\usepackage{url}            
\usepackage{booktabs}       
\usepackage{amsfonts}       
\usepackage{nicefrac}       
\usepackage{microtype}      
\usepackage[table]{xcolor}
\usepackage{wrapfig}
\usepackage{multirow}
\usepackage{multicol}
\usepackage{amsmath}
\usepackage{graphicx}
\usepackage{subfigure}
\usepackage[ruled,linesnumbered]{algorithm2e}

\usepackage{comment}
\usepackage{float}
\usepackage[export]{adjustbox}
\usepackage{caption}

\title{AffectGPT: Dataset and Framework for \\ Explainable Multimodal Emotion Recognition}

\author{
	Zheng Lian$^{1}$, Haiyang Sun$^{1}$, Licai Sun$^{1}$, Jiangyan Yi$^{1}$, Bin Liu$^{1}$, Jianhua Tao$^{2,3}$ \\
	$^1$Institute of Automation, Chinese Academy of Sciences \\
	$^2$Department of Automation, Tsinghua University \\
	$^3$Beijing National Research Center for Information Science and Technology, Tsinghua University \\
	\texttt{lianzheng2016@ia.ac.cn} \\
}

\begin{document}
\maketitle

\begin{abstract}
	Explainable Multimodal Emotion Recognition (EMER) is an emerging task that aims to achieve reliable and accurate emotion recognition. However, due to the high annotation cost, the existing dataset (denoted as EMER-Fine) is small, making it difficult to perform supervised training. To reduce the annotation cost and expand the dataset size, this paper reviews the previous dataset construction process. Then, we simplify the annotation pipeline, avoid manual checks, and replace the closed-source models with open-source models. Finally, we build \textbf{EMER-Coarse}, a coarsely-labeled dataset containing large-scale samples. Besides the dataset, we propose a two-stage training framework \textbf{AffectGPT}. The first stage exploits EMER-Coarse to learn a coarse mapping between multimodal inputs and emotion-related descriptions; the second stage uses EMER-Fine to better align with manually-checked results. Experimental results demonstrate the effectiveness of our proposed method on the challenging EMER task. To facilitate further research, we will make the code and dataset available at: \textcolor[rgb]{0.93,0.0,0.47}{https://github.com/zeroQiaoba/AffectGPT}.
\end{abstract}

\section{Introduction}
Emotion recognition is an important research topic in human-computer interaction. Its main goal is to predict the most likely label from a fixed space \cite{lian2024merbench} (such as the seven basic emotions in Ekman's theory \cite{ekman1970universal}). However, emotions are complex. Limiting the label space and fixing the number of predictions may lead to inaccurate descriptions of emotions. Meanwhile, traditional emotion recognition lacks the explanation process, which is crucial to enhance the annotation reliability.

To this end, researchers propose a new task called Explainable Multimodal Emotion Recognition (EMER) \cite{lian2023explainable}. Unlike traditional emotion recognition, EMER exploits multi-modal and multi-faceted clues to predict emotions in an open-vocabulary (OV) manner. These clues can also serve as support and evidence for these predictions. Therefore, EMER provides a promising way for accurate and reliable emotion recognition. However, due to the high annotation cost, previous works only contain a small number of labeled samples (denoted as EMER-Fine) \cite{lian2023explainable}. These samples can only evaluate the performance of pre-trained systems and are not enough for supervised training.

To reduce the annotation cost, we review the previous dataset construction process. It contains four steps: pre-labeling audio and video clues, manually checking these clues, disambiguating subtitles, and translating to obtain bilingual descriptions \cite{lian2023explainable}. This process relies on manual checks and closed-source models. To reduce the annotation cost, we try to avoid manual checks and use open-source models instead. Then, we build \textbf{EMER-Coarse}, a coarsely-labeled dataset containing large-scale data. Since emotion recognition focuses on identifying human emotional states, we construct this dataset based on MER2024-SEMI \cite{lian2024mer}, which contains 115,595 human-centric videos.

Besides EMER-Coarse, we propose \textbf{AffectGPT}, a two-stage training framework for EMER. In the first stage, we use large-scale EMER-Coarse to learn a coarse alignment between multimodal inputs and emotion-related descriptions. In the second stage, we use small-scale EMER-Fine to better align with manually-checked results. The main contributions of this paper can be summarized as follows:
\begin{itemize}
	\item \textbf{(Dataset)} We build {EMER-Coarse}, a large-scale dataset for EMER. This dataset contains  115,595 samples, much more than previous datasets and sufficient for supervised training.
	
	\item \textbf{(Method)} We propose {AffectGPT}, a two-stage framework for EMER. The first stage learns a coarse mapping and the second stage better aligns with manually-checked results.
	
	\item \textbf{(Performance)} Experimental results demonstrate the effectiveness of this framework. Our systematic analysis can also provide some inspiration for subsequent researchers.
\end{itemize}

\section{Task and Evaluation}
This section reviews the task definition and evaluation metrics of EMER. Unlike traditional emotion recognition, EMER aims to predict emotions in an explainable and open-vocabulary manner. Following previous works \cite{lian2023explainable}, we focus on emotion recognition and use the overlap between predicted and annotated results as the evaluation metric. Since we do not fix the label space, different models may generate synonyms. To remove their impacts, we first group all labels using GPT-3.5 \cite{openai2022chatgpt} (``gpt-3.5-turbo-16k-0613''): \textcolor[rgb]{0.93,0.0,0.47}{\emph{Please assume the role of an expert in the field of emotions. We provide a set of emotions. Please group the emotions, with each group containing emotions with the same meaning. Directly output the results. The output format should be a list containing multiple lists.}}

Specifically, assume that $G(\cdot)$ is the GPT-generated mapping function between labels and group IDs. $\{y_i\}_{i=1}^M$ and $\{\hat{y}_i\}_{i=1}^N$ are the annotated and predicted labels, respectively. Here, $M$ and $N$ are the number of labels. Before metric calculation, we first map each label into its group ID:
\begin{equation}
\mathcal{Y} = \{G(x) |x \in \{y_i\}_{i=1}^M\}, \; \hat{\mathcal{Y}} = \{G(x) |x \in \{\hat{y}_i\}_{i=1}^N\}.
\end{equation}

Then, we calculate the average of precision and recall as the final metric:
\begin{equation}
\mbox{Accuracy}_{\mbox{s}} = \frac{|\mathcal{Y} \cap \hat{\mathcal{Y}}|}{|\hat{\mathcal{Y}}|}, \;\mbox{Recall}_{\mbox{s}} = \frac{|\mathcal{Y} \cap \hat{\mathcal{Y}}|}{|\mathcal{Y}|},
\end{equation}
\begin{equation}
\mbox{Avg} = \frac{\mbox{Accuracy}_{\mbox{s}} + \mbox{Recall}_{\mbox{s}}}{2}.
\end{equation}

\section{EMER-Coarse}
\label{sec:3}
This section reviews the previous dataset construction pipeline \cite{lian2023explainable} and attempts to reduce the annotation cost. Specifically, the previous pipeline consists of four steps: pre-labeling to generate multimodal clues, manual checking these clues, disambiguation of subtitles, and translation to obtain bilingual descriptions. The main cost lies in manual checks and the use of closed-source models for pre-labeling, disambiguation, and translation. To reduce the cost, we try to avoid manual checks and replace these closed-source models with open-source models. In this section, we test the mainstream open-source LLMs and MLLMs. Since the results vary slightly between distinct runs, we run all experiments twice and report the average score and standard deviation.

\subsection{Pre-labeling}
\label{sec:3-1}
Previously, the pre-labeling process relied on the closed-source GPT-4 (``gpt-4-vision-preview''). To find its replacement, we evaluate the performance of some representative open-source MLLMs. According to previous findings \cite{lian2023explainable}, adding subtitles using a two-step strategy can achieve better performance, i.e., first extracting emotion-related descriptions from MLLMs and then using them to disambiguate the subtitle. In this section, we follow this strategy and report results in Table \ref{Table1}. In this table, some results are taken from previous works \cite{lian2023explainable} as they follow the same experimental setup.

Besides the single MLLM, can we achieve better performance if we combine different MLLMs? To answer this question, we further select the top-performing audio and video MLLMs and report the performance of their combinations. In Table \ref{Table1}, we observe that these combinations usually bring performance improvement. Among them, the combination of SALMONN and Chat-UniVi performs best, even surpassing GPT-4. Therefore, we use it for pre-labeling.

\begin{table}[h]
	\centering
	\renewcommand\tabcolsep{2pt}
	\caption{Performance of different MLLMs and their combinations. Following previous works \cite{lian2023explainable}, we consider language influence and report the results under different languages.}
	\label{Table1}
	\scalebox{0.8}{
		\begin{tabular}{lccc|>{\columncolor{lightgray}}ccc|>{\columncolor{lightgray}}ccc}
			\hline
			\multirow{2}{*}{Model} & \multirow{2}{*}{L} & \multirow{2}{*}{V} & \multirow{2}{*}{A} & \multicolumn{3}{c|}{English} & \multicolumn{3}{c}{Chinese} \\
			&&&&Avg & $\mbox{Accuracy}_{\mbox{s}}$ & $\mbox{Recall}_{\mbox{s}}$ & Avg & $\mbox{Accuracy}_{\mbox{s}}$ & $\mbox{Recall}_{\mbox{s}}$\\
			\hline
			\multicolumn{10}{c}{Audio + Subtitle} \\
			\hline
			Qwen-Audio \cite{chu2023qwen}   & $\surd$  & $\times$ & $\surd$&40.23$\pm$0.09 & 49.42$\pm$0.18 & 31.04$\pm$0.00 & 43.53$\pm$0.04 & 53.71$\pm$0.00 & 33.34$\pm$0.09 \\
			OneLLM \cite{han2023onellm}     & $\surd$  & $\times$ & $\surd$&43.04$\pm$0.06 & 45.92$\pm$0.05 & 40.15$\pm$0.06 & 46.77$\pm$0.01 & 52.07$\pm$0.06 & 41.47$\pm$0.08 \\
			SECap \cite{xu2024secap} & $\surd$   & $\times$ & $\surd$&46.94$\pm$0.10 & 54.52$\pm$0.15 & 39.37$\pm$0.05 & 47.09$\pm$0.15 & 55.55$\pm$0.23 & 38.64$\pm$0.08 \\
			SALMONN \cite{tang2023salmonn}      & $\surd$  & $\times$ & $\surd$&48.06$\pm$0.04 & 50.20$\pm$0.04 & 45.92$\pm$0.04 & 48.53$\pm$0.03 & 52.24$\pm$0.00 & 44.82$\pm$0.05 \\
			\hline
			\multicolumn{10}{c}{Video + Subtitle} \\
			\hline
			Otter  \cite{li2023otter}   & $\surd$  & $\surd$ & $\times$&44.40$\pm$0.09 & 50.71$\pm$0.10 & 38.09$\pm$0.09 & 46.92$\pm$0.04 & 52.65$\pm$0.16 & 41.18$\pm$0.08 \\
			VideoChat  \cite{li2023videochat} & $\surd$  & $\surd$ & $\times$&45.70$\pm$0.09 & 42.90$\pm$0.27 & 48.49$\pm$0.10 & 45.63$\pm$0.04 & 47.20$\pm$0.12 & 44.05$\pm$0.05 \\
			Video-LLaMA \cite{zhang2023video}  & $\surd$ & $\surd$ & $\times$&44.74$\pm$0.14 & 44.14$\pm$0.13 & 45.34$\pm$0.15 & 47.27$\pm$0.03 & 47.98$\pm$0.07 & 46.56$\pm$0.01 \\
			Video-LLaVA \cite{lin2023video}   & $\surd$  & $\surd$ & $\times$&47.12$\pm$0.15 & 48.58$\pm$0.02 & 45.66$\pm$0.29 & 49.59$\pm$0.05 & 53.95$\pm$0.03 & 45.23$\pm$0.13 \\
			VideoChat2 \cite{li2024mvbench}   & $\surd$  & $\surd$ & $\times$&49.60$\pm$0.28 & 54.72$\pm$0.41 & 44.47$\pm$0.15 & 49.90$\pm$0.06 & 57.12$\pm$0.08 & 42.68$\pm$0.04 \\
			OneLLM \cite{han2023onellm}     & $\surd$  & $\surd$  & $\times$&50.99$\pm$0.08 & 55.93$\pm$0.09 & 46.06$\pm$0.06 & 51.84$\pm$0.08 & 56.43$\pm$0.04 & 47.26$\pm$0.11 \\
			LLaMA-VID \cite{li2023llama}    & $\surd$  & $\surd$  & $\times$&51.29$\pm$0.09 & 52.71$\pm$0.18 & 49.87$\pm$0.00 & 52.45$\pm$0.02 & 57.30$\pm$0.00 & 47.61$\pm$0.03 \\
			mPLUG-Owl \cite{ye2023mplug}   & $\surd$  & $\surd$ & $\times$&52.79$\pm$0.13 & 54.54$\pm$0.13 & 51.04$\pm$0.13 & 51.43$\pm$0.03 & 56.40$\pm$0.11 & 46.47$\pm$0.18 \\
			Video-ChatGPT \cite{maaz2023video} & $\surd$ & $\surd$ & $\times$&50.73$\pm$0.06 & 54.03$\pm$0.04 & 47.44$\pm$0.07 & 55.34$\pm$0.02 & 61.15$\pm$0.10 & 49.52$\pm$0.06 \\
			Chat-UniVi \cite{jin2023chat}   & $\surd$  & $\surd$  & $\times$&53.09$\pm$0.01 & 53.68$\pm$0.00 & 52.50$\pm$0.02 & 54.20$\pm$0.02 & 58.54$\pm$0.01 & 49.86$\pm$0.03 \\
			GPT-4V \cite{openai2023gpt4v}     & $\surd$   & $\surd$ & $\times$&56.69$\pm$0.04 & 48.52$\pm$0.07 & 64.86$\pm$0.00 & 57.34$\pm$0.01 & 54.61$\pm$0.02 & 60.07$\pm$0.01 \\
			\hline
			\multicolumn{10}{c}{Audio + Video + Subtitle} \\
			\hline
			SECap + mPLUG-Owl & $\surd$ & $\surd$ & $\surd$&57.71$\pm$0.05 & 50.05$\pm$0.23 & 65.38$\pm$0.33 & 55.22$\pm$0.22 & 51.65$\pm$0.27 & 58.79$\pm$0.16 \\
			SALMONN + Video-ChatGPT & $\surd$ & $\surd$ & $\surd$&58.71$\pm$0.24 & 53.16$\pm$0.17 & 64.26$\pm$0.31 & 55.10$\pm$0.16 & 53.44$\pm$0.14 & 56.76$\pm$0.19 \\
			SECap + Video-ChatGPT & $\surd$ & $\surd$ & $\surd$&57.41$\pm$0.09 & 52.03$\pm$0.04 & 62.79$\pm$0.14 & 56.49$\pm$0.02 & 56.50$\pm$0.01 & 56.48$\pm$0.05 \\
			SECap + Chat-UniVi & $\surd$ & $\surd$ & $\surd$&59.13$\pm$0.08 & 48.85$\pm$0.29 & 69.41$\pm$0.13 & 56.49$\pm$0.14 & 52.38$\pm$0.07 & 60.59$\pm$0.22 \\
			SALMONN + mPLUG-Owl & $\surd$ & $\surd$ & $\surd$&59.77$\pm$0.05 & 51.77$\pm$0.01 & 67.76$\pm$0.11 & 55.94$\pm$0.21 & 51.74$\pm$0.19 & 60.14$\pm$0.23 \\
			SALMONN + Chat-UniVi & $\surd$ & $\surd$ & $\surd$&59.47$\pm$0.08 & 51.62$\pm$0.00 & 67.31$\pm$0.15 & 57.54$\pm$0.06 & 51.65$\pm$0.06 & 63.42$\pm$0.06 \\
			\hline
			EMER(Multi) & $\surd$  & $\surd$  & $\surd$&80.05$\pm$0.24 & 80.03$\pm$0.37 & 80.07$\pm$0.10 & 85.20$\pm$0.03 & 87.09$\pm$0.00 & 83.31$\pm$0.05 \\
			\hline
		\end{tabular}
	}
\end{table}

\subsection{Disambiguation and Translation}
Disambiguation and translation deal with plain text data and these modules previously relied on GPT-3.5. To find its alternative, we test some typical open-source LLMs. Experimental results are shown in Table \ref{Table2}. We observe that if only the translation module is replaced with open-source LLMs, the performance drop is small. But if we replace both translation and disambiguation, the performance drop is obvious. These results show that for non-complex tasks (e.g., translation), the performance of open-source LLMs is close to GPT-3.5. But for complex tasks (e.g., disambiguation), there is still a gap between open-source LLMs and GPT-3.5. The reason may be that we do not test larger LLMs due to limited GPU memory. Generally, larger LLMs help solve more complex tasks, which is left for our future work. Meanwhile, we observe that Qwen2-7B performs better than LLaMA3-8B in translation. Therefore, we use Qwen2-7B for translation and GPT-3.5 for disambiguation. This replacement reduces the OpenAI API call cost and maintains the overall performance.

Finally, we use the above strategy to automatically annotate MER2024-SEMI \cite{lian2024mer}. These annotation results take into account all acoustic, visual, and lexical clues. Since these results have not been manually checked, there may be some inaccuracies. We call this dataset \textbf{EMER-Coarse}.

\begin{table}[h]
	\centering
	\caption{Choice of open-source LLMs for translation and disambiguation. Since the combination of SALMONN and Chat-UniVi performs best (see Table \ref{Table1}), we conduct analysis on this combination.}
	\label{Table2}
	\scalebox{0.8}{
		\begin{tabular}{lc|>{\columncolor{lightgray}}ccc|>{\columncolor{lightgray}}ccc}
			\hline
			\multirow{2}{*}{Translate} & \multirow{2}{*}{Disambiguate} & \multicolumn{3}{c|}{English} & \multicolumn{3}{c}{Chinese} \\
			&&Avg & $\mbox{Accuracy}_{\mbox{s}}$ & $\mbox{Recall}_{\mbox{s}}$ & Avg & $\mbox{Accuracy}_{\mbox{s}}$ & $\mbox{Recall}_{\mbox{s}}$\\
			\hline
			
			GPT-3.5 & GPT-3.5 &59.47$\pm$0.08 & 51.62$\pm$0.00 & 67.31$\pm$0.15 & 57.54$\pm$0.06 & 51.65$\pm$0.06 & 63.42$\pm$0.06 \\
			
			\hline
			LLaMA3-8B  & GPT-3.5 &57.13$\pm$0.27 & 49.63$\pm$0.32 & 64.64$\pm$0.22 & 55.50$\pm$0.02 & 50.85$\pm$0.19 & 60.15$\pm$0.16 \\
			LLaMA3-8B & LLaMA3-8B &55.50$\pm$0.09 & 49.91$\pm$0.04 & 61.08$\pm$0.22 & 52.59$\pm$0.74 & 47.03$\pm$0.42 & 58.15$\pm$1.05 \\
			\hline
			Qwen2-7B  & GPT-3.5  &58.22$\pm$0.11 & 49.68$\pm$0.21 & 66.76$\pm$0.00 & 56.65$\pm$0.27 & 52.95$\pm$0.23 & 60.36$\pm$0.32 \\
			Qwen2-7B &Qwen2-7B &53.38$\pm$0.60 & 44.74$\pm$0.67 & 62.01$\pm$0.54 & 55.15$\pm$0.03 & 47.92$\pm$0.06 & 62.37$\pm$0.12 \\
			\hline
		\end{tabular}
	}
\end{table}

\section{AffectGPT}
Besides EMER-Coarse, we propose a two-stage framework AffectGPT. This section introduces this framework from three aspects: training process, model architecture, and experimental setup.

\paragraph{Training Process}
The first stage uses EMER-Coarse to learn a coarse alignment between multimodal inputs and emotion-related outputs. The second stage uses EMER-Fine to better align with manually-checked results. Considering that EMER-Fine has more reliable labels, we evaluate the performance of different systems on it. However, the second stage is also trained on EMER-Fine, so we further split it into training and test sets. The statistics are shown in Table \ref{Table3}.

\begin{wraptable}{r}{5cm}
	\caption{Dataset statistics.}
	\label{Table3}
	\scalebox{0.8}{
		\begin{tabular}{l|c|c}
			\hline
			Dataset & Split & \# of samples \\
			\hline
			EMER-Coarse &-- & {115,595} \\
			\hline
			\multirow{3}{*}{EMER-Fine} &train & 266 \\
			&test  & 66  \\
			&whole & 332 \\
			\hline
		\end{tabular}
	}
\end{wraptable}

\paragraph{Model Architecture}
AffectGPT is borrowed from Video-LLaMA with some modifications. Considering that the origin framework trains audio and video branches separately but emotion recognition requires the integration of multimodal clues, we modify Video-LLaMA to support audio-video-text alignment training. Specifically, we input the audio, video, and subtitle simultaneously, and try to learn a mapping between multimodal inputs and emotion-related descriptions. The reason why we do not design more effective frameworks but use Video-LLaMA is that the main purpose of this paper is to study the effectiveness of EMER-Coarse and the two-stage training process. The impact of different model architectures is left to our future work.

\paragraph{Experimental Setup}
AffectGPT is implemented with PyTorch. All training and inference processes are carried out with an 80G NVIDIA Tesla A100 GPU. During training, we set the maximum number of epochs to 100. Due to the different number of training samples in each stage, the first stage iterates 1000 times per epoch and the second stage iterates 88 times per epoch. Meanwhile, we set the batch size of each iteration to 3. Limited by our GPU memory capacity, we do not test a larger batch size. During training, we freeze the weights of the acoustic encoder, visual encoder, and LLM, and only train Q-Former to learn the mapping between unimodal encoders and LLM.

\section{Results and Discussion}
\label{sec:5}
AffectGPT is a two-stage training framework. To verify its effectiveness, we perform ablation studies on each stage. Considering that Video-LLaMA provides pretrained Q-Formers, we first reveal their necessity and study whether AffectGPT can be trained directly on randomly initialized weights. Then, we study the impact of different LLMs and discuss the necessity of each stage. Finally, we show the performance of AffectGPT on the EMER task. For convenience, in this section, we abbreviate the first stage as \emph{stage1} and the second stage as \emph{stage2}.

During training, AffectGPT learns a mapping between audio-video-text inputs and emotion-related outputs. These outputs are in English and have already considered the disambiguation process (see Section \ref{sec:3}). In the previous evaluation pipeline (see Table \ref{Table1}), we need additional translation and disambiguation operations, which increases the evaluation cost. To reduce the cost, in this section, we extract emotion labels directly from the output of AffectGPT for performance evaluation.

\subsection{Ablation Study on Stage1}

\begin{figure*}[h]
	\begin{center}
		\subfigure[\scriptsize{Pretrain + EMER-Fine(Whole)}]{
			\label{Figure1-1}
			\centering
			\includegraphics[width=0.316\linewidth, trim=0 0 0 0]{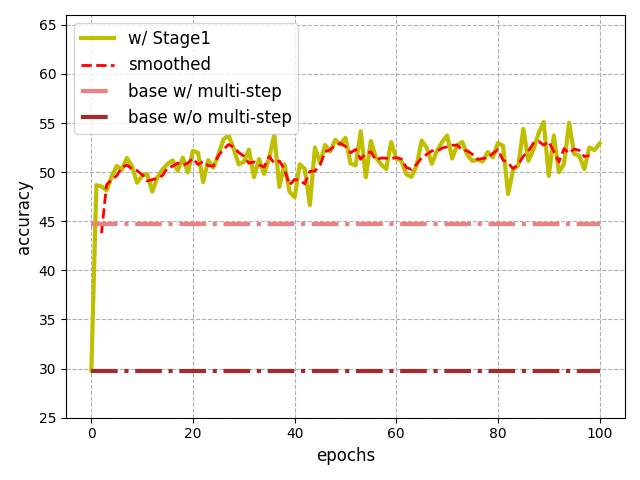}
		}
		\subfigure[\scriptsize{Pretrain + EMER-Fine(Train)}]{
			\label{Figure1-2}
			\centering
			\includegraphics[width=0.316\linewidth, trim=0 0 0 0]{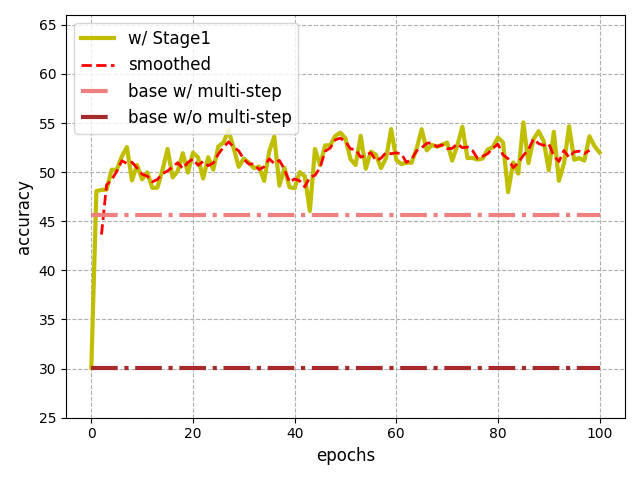}
		}
		\subfigure[\scriptsize{Pretrain + EMER-Fine(Test)}]{
			\label{Figure1-3}
			\centering
			\includegraphics[width=0.316\linewidth, trim=0 0 0 0]{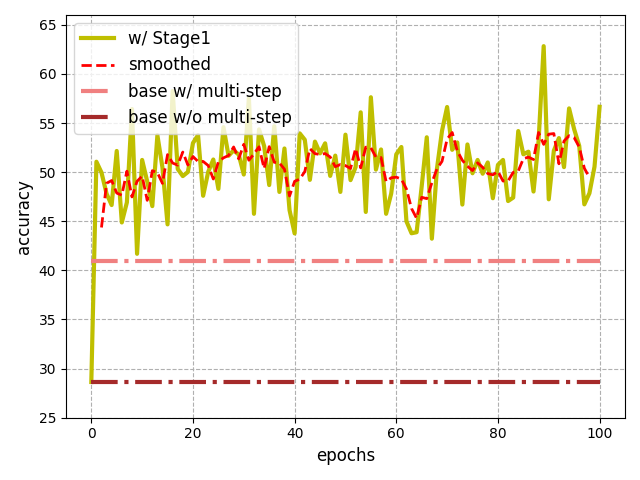}
		}
		\subfigure[\scriptsize{Random + EMER-Fine(Whole)}]{
			\label{Figure1-4}
			\centering
			\includegraphics[width=0.316\linewidth, trim=0 0 0 0]{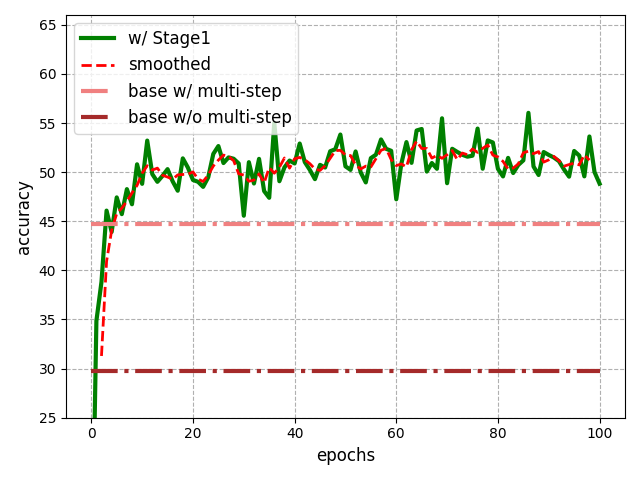}
		}
		\subfigure[\scriptsize{Random + EMER-Fine(Train)}]{
			\label{Figure1-5}
			\centering
			\includegraphics[width=0.316\linewidth, trim=0 0 0 0]{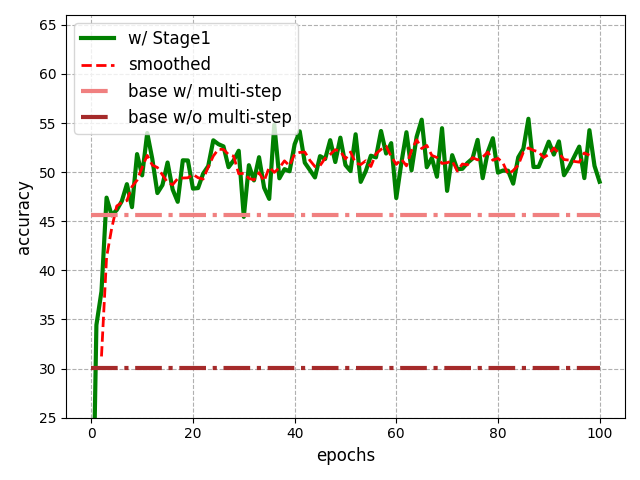}
		}
		\subfigure[\scriptsize{Random + EMER-Fine(Test)}]{
			\label{Figure1-6}
			\centering
			\includegraphics[width=0.316\linewidth, trim=0 0 0 0]{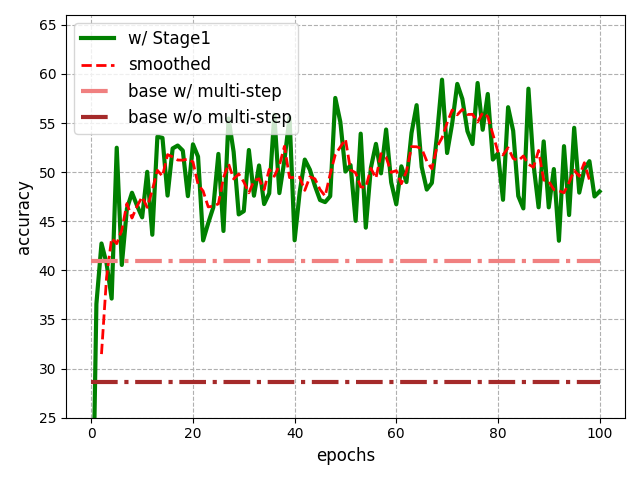}
		}
	\end{center}
	\caption{Ablation study on \emph{stage1}. In these figures, we train models with different initialization strategies and report their results on different sets. Besides the original accuracy curve, we also add a smoothed curve. Meanwhile, we introduce two baselines without \emph{stage1} training.}
	\label{Figure1}
\end{figure*}

\paragraph{Choice of Evaluation Set}
Video-LLaMA provides pretrained Q-Formers. In this section, we try to analyze whether these weights can help the model converge and achieve better performance. Before comparing different initialization strategies, we need to determine which dataset should be used for evaluation. In this paper, we have three choices: the training set, the test set, and the entire EMER-Fine. In Figure \ref{Figure1}, we present the results on different datasets. We observe that increasing the number of samples can reduce the fluctuation of accuracy and help us draw more reliable conclusions. Therefore, in \emph{stage1}, we evaluate the performance on the entire EMER-Fine. It should be noted that further increasing the dataset size may obtain more stable results, therefore we plan to expand EMER-Fine in the future.

\begin{figure*}[h]
	\begin{center}
		\subfigure[Training loss]{
			\label{Figure2-1}
			\centering
			\includegraphics[width=0.39\linewidth, trim=0 0 0 0]{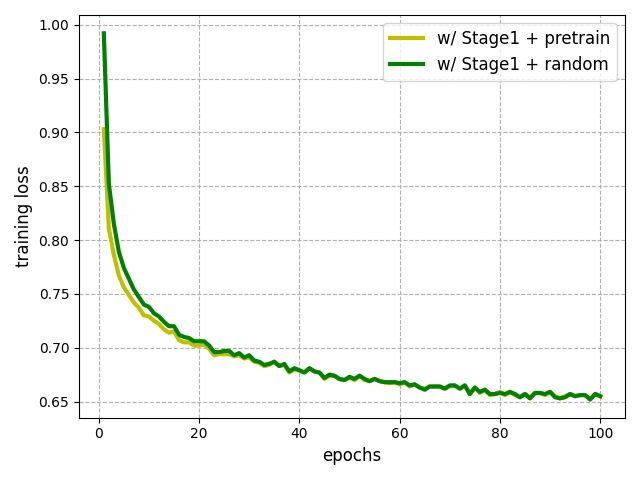}
		}
		\subfigure[Accuracy]{
			\label{Figure2-2}
			\centering
			\includegraphics[width=0.39\linewidth, trim=0 0 0 0]{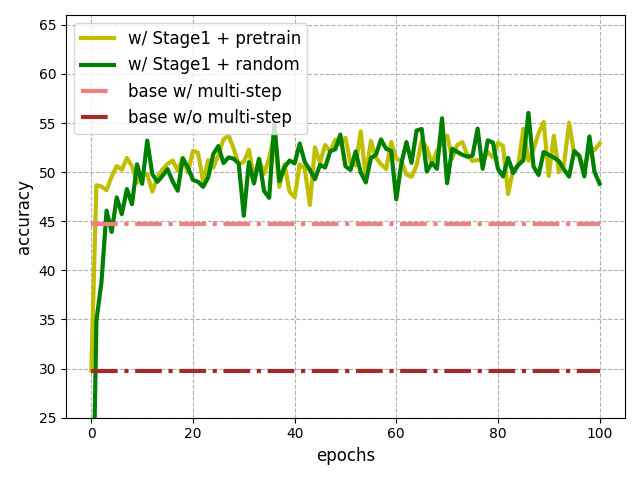}
		}
	\end{center}
	\caption{Impact of different initialization strategies. We plot the curve of training loss and accuracy. As for accuracy, we evaluate the performance on the entire EMER-Fine.}
	\label{Figure2}
\end{figure*}

\paragraph{Impart of Initialization Strategies}
Figure \ref{Figure2} reveals the impact of different initialization strategies. Figure \ref{Figure2-1} shows the curve of training loss. We observe that the model converges around 100 epochs, which proves the rationality of our choice of the maximum number of epochs. Meanwhile, different initialization strategies only have impacts in the initial epochs, and the model will eventually converge to a similar loss. Figure \ref{Figure2-2} shows the emotion recognition results. We observe that different initialization strategies have limited impacts, proving that our large-scale EMER-Coarse is sufficient to train the model from randomly initialized weights. Therefore, we can conclude that the initialization strategy has limited impact in \emph{stage1} training.

\begin{figure*}[h]
	\begin{center}
		\subfigure[Training loss]{
			\label{Figure3-1}
			\centering
			\includegraphics[width=0.39\linewidth, trim=0 0 0 0]{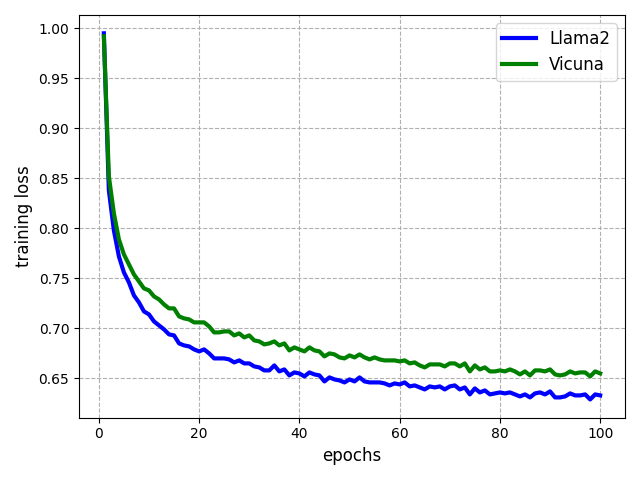}
		}
		\subfigure[Accuracy]{
			\label{Figure3-2}
			\centering
			\includegraphics[width=0.39\linewidth, trim=0 0 0 0]{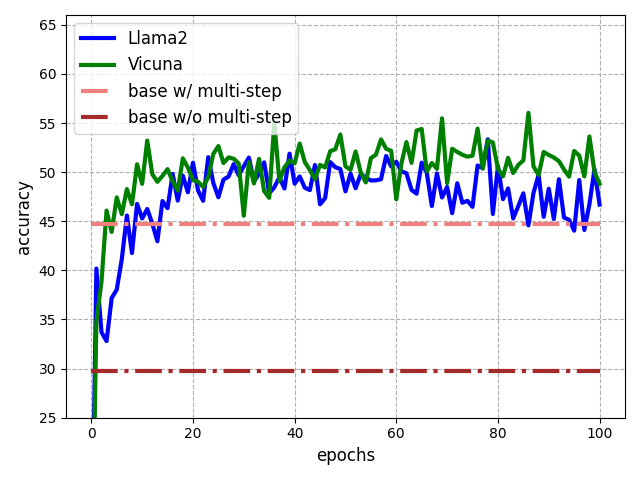}
		}
	\end{center}
	\caption{Impact of different LLMs. We use the random initialization strategy and evaluate the performance on the entire EMER-Fine.}
	\label{Figure3}
\end{figure*}

\paragraph{Choice of LLMs}
This section analyzes the impact of different LLMs. The original Video-LLaMA uses Vicuna (a model based on LLaMA). We try to replace Vicuna with LLaMA-2-Chat (a model based on LLaMA-2) and study its impact. The pretrained Q-Former provided by Video-LLaMA is only used to connect encoders and Vicuna. If we replace the LLM, we cannot use these pretrained weights. For a fair comparison, all experiments adopt the random initialization strategy, and experimental results are shown in Figure\ref{Figure3}. Figure \ref{Figure3-1} shows the training loss and Figure \ref{Figure3-2} shows the emotion recognition results. Interestingly, we observe that the training loss of LLaMA-2 is lower than that of Vicuna, but Vicuna performs better than LLaMA-2 in emotion recognition. The reason may be that we fix the weights of LLMs and do not use LoRA for supervised fine-tuning, which may limit the performance of LLaMA-2 on downstream tasks. Meanwhile, these results also prove that there is no strong correlation between training loss and test accuracy. From another perspective, these results also show that LLMs affect the performance of AffectGPT. Therefore, we plan to explore the impact of other LLMs in the future.

\paragraph{Effectiveness of Stage1}
In Figures \ref{Figure1}$\sim$\ref{Figure3}, we add two baselines, both of which rely on Video-LLaMA. Specifically, one uses a multi-step strategy, i.e., first extracts emotion-related descriptions from Video-LLaMA and then uses these descriptions to disambiguate subtitles. The other does not use a multi-step strategy, i.e., directly inputs audio-video-text clues into Video-LLaMA. From Figures \ref{Figure1}$\sim$\ref{Figure3}, we can see that no matter which initialization strategy and which LLM are used, our AffectGPT always outperforms two baselines. These results demonstrate the effectiveness of \emph{stage1}. That is, training on EMER-Coarse usually leads to performance improvements.




\subsection{Ablation Study on Stage2}

\begin{figure*}[h]
	\begin{center}
		\subfigure[\scriptsize{EMER-Fine(Whole)}]{
			\label{Figure4-1}
			\centering
			\includegraphics[width=0.316\linewidth, trim=0 0 0 0]{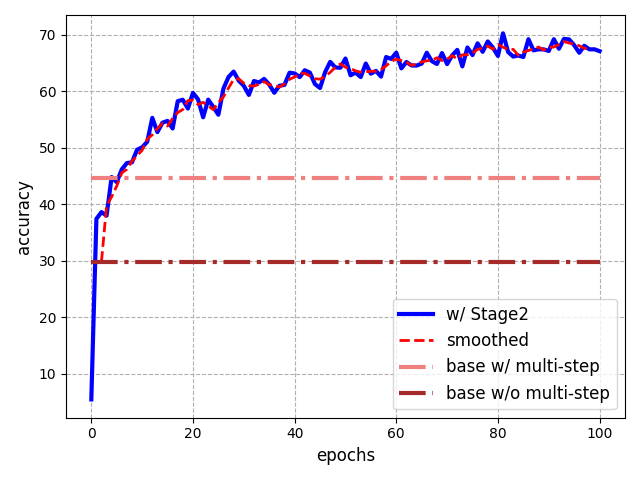}
		}
		\subfigure[\scriptsize{EMER-Fine(Train)}]{
			\label{Figure4-2}
			\centering
			\includegraphics[width=0.316\linewidth, trim=0 0 0 0]{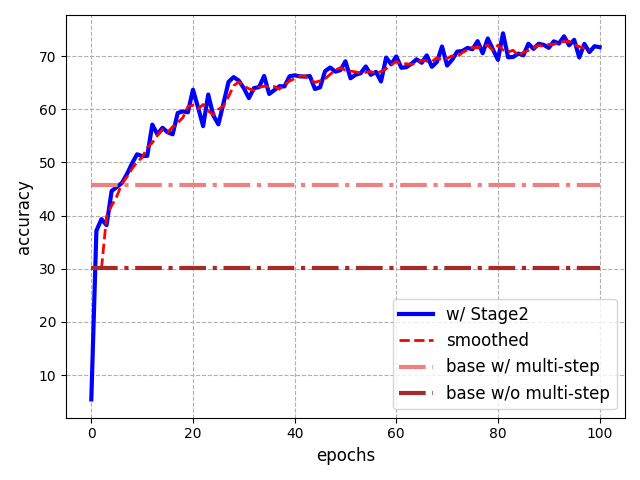}
		}
		\subfigure[\scriptsize{EMER-Fine(Test)}]{
			\label{Figure4-3}
			\centering
			\includegraphics[width=0.316\linewidth, trim=0 0 0 0]{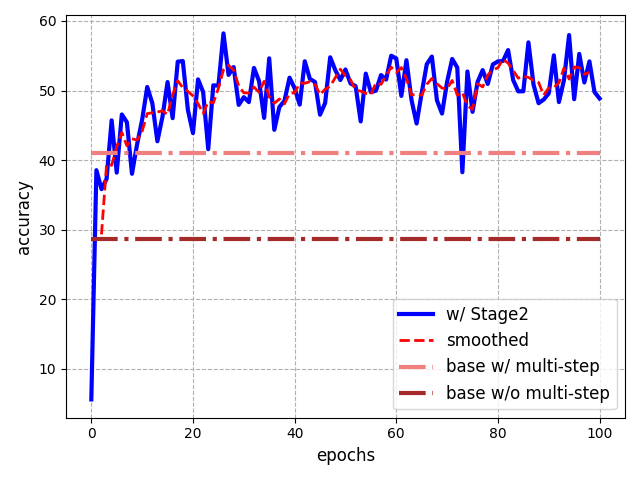}
		}
	\end{center}
	\caption{Ablation study on stage2. In these figures, we show the results on different subsets.}
	\label{Figure4}
\end{figure*}

\paragraph{Choice of Evaluation Set} 
In \emph{stage1}, we choose the entire EMER-Fine for performance evaluation. But for \emph{stage2}, which part of the dataset should we use? Figure \ref{Figure4} shows the results on different sets. In Figure \ref{Figure4-2}, we observe that the training accuracy steadily improves with increasing epochs. These results prove that our model can well fit training data. It is not appropriate to use the training accuracy for performance evaluation. In Figure \ref{Figure4-3}, we observe that the test accuracy fluctuates greatly. The reason may be that the test data is limited. Therefore, in subsequent analysis, we use the smoothed test accuracy for performance evaluation.

\begin{figure*}[h]
	\begin{center}
		\subfigure[Training loss (pretrain)]{
			\label{Figure5-1}
			\centering
			\includegraphics[width=0.39\linewidth, trim=0 0 0 0]{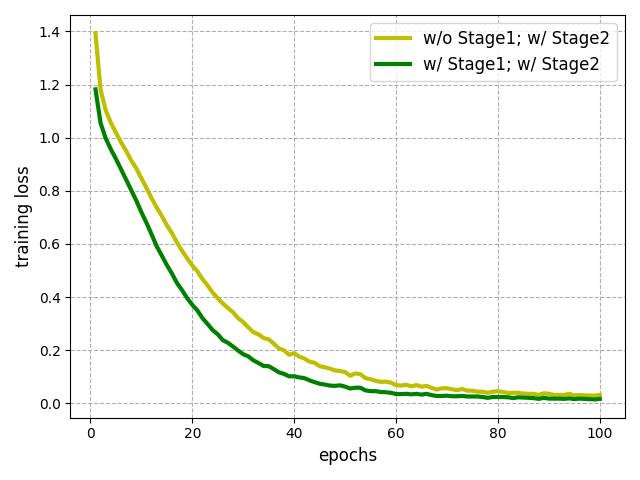}
		}
		\subfigure[Accuracy (pretrain)]{
			\label{Figure5-2}
			\centering
			\includegraphics[width=0.39\linewidth, trim=0 0 0 0]{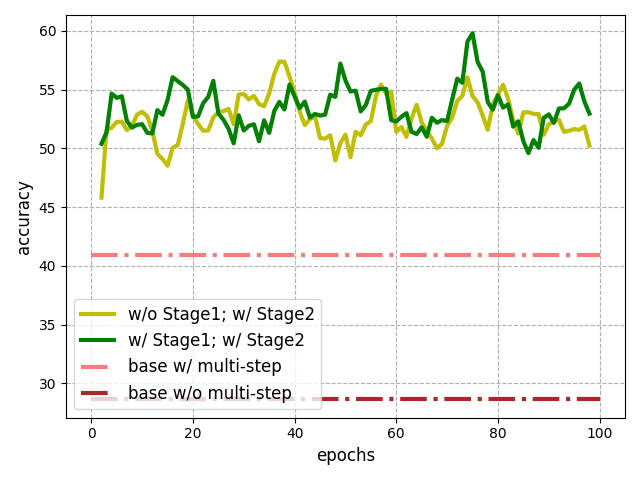}
		}
		
		\subfigure[Training loss (random)]{
			\label{Figure5-3}
			\centering
			\includegraphics[width=0.39\linewidth, trim=0 0 0 0]{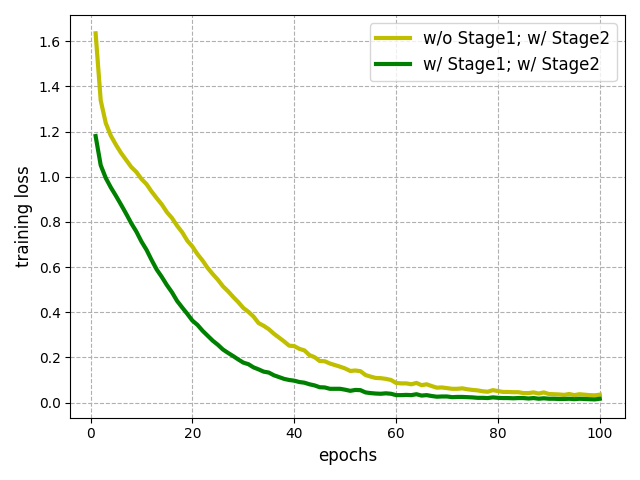}
		}
		\subfigure[Accuracy (random)]{
			\label{Figure5-4}
			\centering
			\includegraphics[width=0.39\linewidth, trim=0 0 0 0]{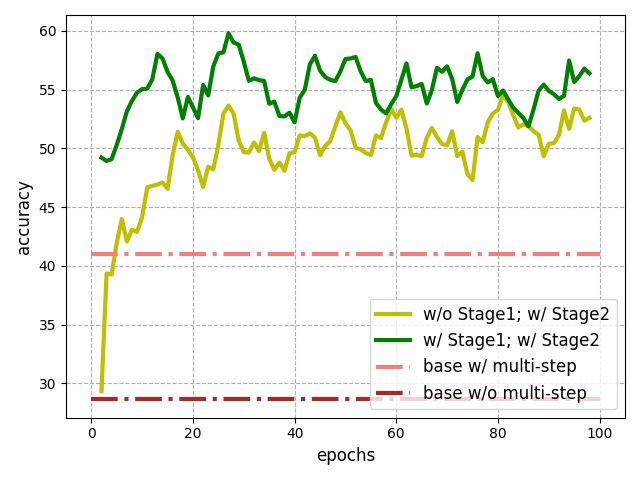}
		}
	\end{center}
	\caption{Necessity of two-stage training framework.}
	\label{Figure5}
\end{figure*}

\paragraph{Necessity of Two-stage Training}
AffectGPT is a two-stage training framework. But can we only train on \emph{stage2} and ignore \emph{stage1}? This section attempts to study the necessity of each stage under different initialization strategies. Experimental results are shown in Figure \ref{Figure5}. From the training loss (see Figures \ref{Figure5-1} and \ref{Figure5-3}), we observe that with the help of \emph{stage1}, the model can obtain better initialization weights, so that it converges faster during \emph{stage2}. From the test accuracy (see Figures \ref{Figure5-2} and \ref{Figure5-4}), we observe that the model with \emph{stage1} usually performs better than the models without \emph{stage1}. This phenomenon is more obvious under the random initialization strategy. From another perspective, we cannot ignore \emph{stage1} and use the random initialization strategy at the same time. Limited EMER-Fine is not enough to train a well-performing model from scratch.

\subsection{Main Results}
In Table \ref{Table4}, we show the performance of AffectGPT on the test set of EMER-Fine under different initialization strategies. Compared with the original Video-LLaMA (w/o \emph{stage1} and w/o \emph{stage2}), training on EMER-Coarse and EMER-Fine remarkably improves the performance. These results reveal the quality of our EMER-Coarse dataset. Meanwhile, two-stage results are generally better than one-stage results, which further demonstrates the effectiveness of our two-stage framework.

\begin{table}[h]
	\centering
	\caption{Performance of AffectGPT. We report results on the test set of EMER-Fine.}
	\label{Table4}
	\scalebox{0.8}{
		\begin{tabular}{cc|ccc|ccc}
			\hline
			\multirow{2}{*}{Stage1} & \multirow{2}{*}{Stage2} & \multicolumn{3}{c|}{Pretrained Weights} & \multicolumn{3}{c}{Random Weights} \\
			& &Avg & $\mbox{Accuracy}_{\mbox{s}}$ & $\mbox{Recall}_{\mbox{s}}$ &Avg & $\mbox{Accuracy}_{\mbox{s}}$ & $\mbox{Recall}_{\mbox{s}}$ \\
			\hline
			-- & --   & 28.64  & 32.22  & 25.05  & 05.87  & 07.58  & 04.17    \\
			\rowcolor{lightgray}
			-- & best   & 61.75  & 62.03  & 61.46 & 58.22  & 59.60  & 56.84 \\
			\hline
			50-epoch & --  & 53.82  & 48.04  & 59.60   & 50.06  & 42.36  & 57.76  \\
			\rowcolor{lightgray}
			50-epoch & best   & 62.78  & 63.11  & 62.45 & 65.08  & 64.29  & 65.86 \\
			\hline
			100-epoch & --   & 56.65  & 47.53  & 65.78  & 48.04  & 40.51  & 55.56  \\
			\rowcolor{lightgray}
			100-epoch & best  & 64.56  & 64.49  & 64.62 & 62.88  & 65.91  & 59.85 \\
			\hline
		\end{tabular}
	}
\end{table}

\section{Conclusion}
\label{sec:6}
EMER is a newly proposed task that aims to achieve reliable and accurate emotion recognition. To promote its development, we propose EMER-Coarse (a large-scale coarsely-labeled dataset) and AffectGPT (a two-stage training framework). Meanwhile, we reveal the impact of each module and study the influence of different initialization strategies and LLMs. Overall, this paper can serve as a complement to existing works on EMER. 


\bibliography{mybib}
\bibliographystyle{unsrt}

\clearpage
\appendix

\section{Details about MLLMs}
\label{appendix_sec:mllm}
In Table \ref{Table5}, we provide model cards for different MLLMs.

\begin{table*}[h]
	\centering
	\renewcommand\arraystretch{1.06}
	\caption{Model cards for MLLMs.}
	\label{Table5}
	\scalebox{0.8}{
		\begin{tabular}{l|l|l}
			\hline
			Models & Support Modality &Link \\
			\hline
			Otter & Video, Text & \textcolor[rgb]{0.93,0.0,0.47}{https://github.com/Luodian/Otter} \\
		
			VideoChat & Video, Text & \textcolor[rgb]{0.93,0.0,0.47}{https://github.com/OpenGVLab/Ask-Anything/tree/main/video\_chat} \\
			VideoChat2 & Video, Text & \textcolor[rgb]{0.93,0.0,0.47}{https://github.com/OpenGVLab/Ask-Anything/tree/main/video\_chat2} \\
			Video-LLaVA & Video, Text & \textcolor[rgb]{0.93,0.0,0.47}{https://github.com/PKU-YuanGroup/Video-LLaVA} \\
			Video-LLaMA & Video, Text & \textcolor[rgb]{0.93,0.0,0.47}{https://github.com/DAMO-NLP-SG/Video-LLaMA} \\
			Video-ChatGPT & Video, Text & \textcolor[rgb]{0.93,0.0,0.47}{https://github.com/mbzuai-oryx/Video-ChatGPT} \\
			LLaMA-VID & Video, Text & \textcolor[rgb]{0.93,0.0,0.47}{https://github.com/dvlab-research/LLaMA-VID} \\
			mPLUG-Owl & Video, Text & \textcolor[rgb]{0.93,0.0,0.47}{https://github.com/X-PLUG/mPLUG-Owl} \\
			Chat-UniVi & Video, Text & \textcolor[rgb]{0.93,0.0,0.47}{https://github.com/PKU-YuanGroup/Chat-UniVi} \\
			\hline
			SALMONN & Audio, Text & \textcolor[rgb]{0.93,0.0,0.47}{https://github.com/bytedance/SALMONN} \\
			Qwen-Audio & Audio, Text & \textcolor[rgb]{0.93,0.0,0.47}{https://github.com/QwenLM/Qwen-Audio} \\
			SECap & Audio, Text & \textcolor[rgb]{0.93,0.0,0.47}{https://github.com/thuhcsi/SECap} \\
			\hline
			OneLLM & Audio, Video, Text & \textcolor[rgb]{0.93,0.0,0.47}{https://github.com/csuhan/OneLLM} \\
			PandaGPT & Audio, Video, Text & \textcolor[rgb]{0.93,0.0,0.47}{https://github.com/yxuansu/PandaGPT} \\
			\hline
			
		\end{tabular}
	}
\end{table*}

\section{Baseline Results on Different Subsets}
In Tables \ref{Table6}$\sim$\ref{Table8}, we report the results of different MLLMs on three subsets of EMER-Fine.

\begin{table}[h]
	\centering
	\renewcommand\tabcolsep{2pt}
	\caption{Main results on the entire EMER-Fine (332 samples).}
	\label{Table6}
	\scalebox{0.8}{
		\begin{tabular}{lccc|>{\columncolor{lightgray}}ccc|>{\columncolor{lightgray}}ccc}
			\hline
			\multirow{2}{*}{Model} & \multirow{2}{*}{L} & \multirow{2}{*}{V} & \multirow{2}{*}{A} & \multicolumn{3}{c|}{English} & \multicolumn{3}{c}{Chinese} \\
			&&&&Avg & $\mbox{Accuracy}_{\mbox{s}}$ & $\mbox{Recall}_{\mbox{s}}$ & Avg & $\mbox{Accuracy}_{\mbox{s}}$ & $\mbox{Recall}_{\mbox{s}}$\\
			\hline
			\multicolumn{10}{c}{Audio + Subtitle} \\
			\hline
			Qwen-Audio \cite{chu2023qwen}   & $\surd$  & $\times$ & $\surd$&40.23$\pm$0.09 & 49.42$\pm$0.18 & 31.04$\pm$0.00 & 43.53$\pm$0.04 & 53.71$\pm$0.00 & 33.34$\pm$0.09 \\
			OneLLM \cite{han2023onellm}     & $\surd$  & $\times$ & $\surd$&43.04$\pm$0.06 & 45.92$\pm$0.05 & 40.15$\pm$0.06 & 46.77$\pm$0.01 & 52.07$\pm$0.06 & 41.47$\pm$0.08 \\
			SECap \cite{xu2024secap} & $\surd$   & $\times$ & $\surd$&46.94$\pm$0.10 & 54.52$\pm$0.15 & 39.37$\pm$0.05 & 47.09$\pm$0.15 & 55.55$\pm$0.23 & 38.64$\pm$0.08 \\
			SALMONN \cite{tang2023salmonn}      & $\surd$  & $\times$ & $\surd$&48.06$\pm$0.04 & 50.20$\pm$0.04 & 45.92$\pm$0.04 & 48.53$\pm$0.03 & 52.24$\pm$0.00 & 44.82$\pm$0.05 \\
			\hline
			\multicolumn{10}{c}{Video + Subtitle} \\
			\hline
			Otter  \cite{li2023otter}   & $\surd$  & $\surd$ & $\times$&44.40$\pm$0.09 & 50.71$\pm$0.10 & 38.09$\pm$0.09 & 46.92$\pm$0.04 & 52.65$\pm$0.16 & 41.18$\pm$0.08 \\
			VideoChat  \cite{li2023videochat} & $\surd$  & $\surd$ & $\times$&45.70$\pm$0.09 & 42.90$\pm$0.27 & 48.49$\pm$0.10 & 45.63$\pm$0.04 & 47.20$\pm$0.12 & 44.05$\pm$0.05 \\
			{Video-LLaMA} \cite{zhang2023video}  & $\surd$ & $\surd$ & $\times$& {44.74$\pm$0.14} & 44.14$\pm$0.13 & 45.34$\pm$0.15 & 47.27$\pm$0.03 & 47.98$\pm$0.07 & 46.56$\pm$0.01 \\
			Video-LLaVA \cite{lin2023video}   & $\surd$  & $\surd$ & $\times$&47.12$\pm$0.15 & 48.58$\pm$0.02 & 45.66$\pm$0.29 & 49.59$\pm$0.05 & 53.95$\pm$0.03 & 45.23$\pm$0.13 \\
			VideoChat2 \cite{li2024mvbench}   & $\surd$  & $\surd$ & $\times$&49.60$\pm$0.28 & 54.72$\pm$0.41 & 44.47$\pm$0.15 & 49.90$\pm$0.06 & 57.12$\pm$0.08 & 42.68$\pm$0.04 \\
			OneLLM \cite{han2023onellm}     & $\surd$  & $\surd$  & $\times$&50.99$\pm$0.08 & 55.93$\pm$0.09 & 46.06$\pm$0.06 & 51.84$\pm$0.08 & 56.43$\pm$0.04 & 47.26$\pm$0.11 \\
			LLaMA-VID \cite{li2023llama}    & $\surd$  & $\surd$  & $\times$&51.29$\pm$0.09 & 52.71$\pm$0.18 & 49.87$\pm$0.00 & 52.45$\pm$0.02 & 57.30$\pm$0.00 & 47.61$\pm$0.03 \\
			mPLUG-Owl \cite{ye2023mplug}   & $\surd$  & $\surd$ & $\times$&52.79$\pm$0.13 & 54.54$\pm$0.13 & 51.04$\pm$0.13 & 51.43$\pm$0.03 & 56.40$\pm$0.11 & 46.47$\pm$0.18 \\
			Video-ChatGPT \cite{maaz2023video} & $\surd$ & $\surd$ & $\times$&50.73$\pm$0.06 & 54.03$\pm$0.04 & 47.44$\pm$0.07 & 55.34$\pm$0.02 & 61.15$\pm$0.10 & 49.52$\pm$0.06 \\
			Chat-UniVi \cite{jin2023chat}   & $\surd$  & $\surd$  & $\times$&53.09$\pm$0.01 & 53.68$\pm$0.00 & 52.50$\pm$0.02 & 54.20$\pm$0.02 & 58.54$\pm$0.01 & 49.86$\pm$0.03 \\
			\hline
			\multicolumn{10}{c}{Audio + Video + Subtitle} \\
			\hline
			SECap + mPLUG-Owl & $\surd$ & $\surd$ & $\surd$&57.71$\pm$0.05 & 50.05$\pm$0.23 & 65.38$\pm$0.33 & 55.22$\pm$0.22 & 51.65$\pm$0.27 & 58.79$\pm$0.16 \\
			SALMONN + Video-ChatGPT & $\surd$ & $\surd$ & $\surd$&58.71$\pm$0.24 & 53.16$\pm$0.17 & 64.26$\pm$0.31 & 55.10$\pm$0.16 & 53.44$\pm$0.14 & 56.76$\pm$0.19 \\
			SECap + Video-ChatGPT & $\surd$ & $\surd$ & $\surd$&57.41$\pm$0.09 & 52.03$\pm$0.04 & 62.79$\pm$0.14 & 56.49$\pm$0.02 & 56.50$\pm$0.01 & 56.48$\pm$0.05 \\
			SECap + Chat-UniVi & $\surd$ & $\surd$ & $\surd$&59.13$\pm$0.08 & 48.85$\pm$0.29 & 69.41$\pm$0.13 & 56.49$\pm$0.14 & 52.38$\pm$0.07 & 60.59$\pm$0.22 \\
			SALMONN + mPLUG-Owl & $\surd$ & $\surd$ & $\surd$&59.77$\pm$0.05 & 51.77$\pm$0.01 & 67.76$\pm$0.11 & 55.94$\pm$0.21 & 51.74$\pm$0.19 & 60.14$\pm$0.23 \\
			SALMONN + Chat-UniVi & $\surd$ & $\surd$ & $\surd$&{59.47}$\pm$0.08 & 51.62$\pm$0.00 & 67.31$\pm$0.15 & 57.54$\pm$0.06 & 51.65$\pm$0.06 & 63.42$\pm$0.06 \\
			\hline
			EMER(Multi) & $\surd$  & $\surd$  & $\surd$&80.05$\pm$0.24 & 80.03$\pm$0.37 & 80.07$\pm$0.10 & 85.20$\pm$0.03 & 87.09$\pm$0.00 & 83.31$\pm$0.05 \\
			\hline
		\end{tabular}
	}
\end{table}

\begin{table}[h]
	\centering
	\renewcommand\tabcolsep{2pt}
	\caption{Main results on the training set of EMER-Fine (266 samples).}
	\label{Table7}
	\scalebox{0.8}{
		\begin{tabular}{lccc|>{\columncolor{lightgray}}ccc|>{\columncolor{lightgray}}ccc}
			\hline
			\multirow{2}{*}{Model} & \multirow{2}{*}{L} & \multirow{2}{*}{V} & \multirow{2}{*}{A} & \multicolumn{3}{c|}{English} & \multicolumn{3}{c}{Chinese} \\
			&&&&Avg & $\mbox{Accuracy}_{\mbox{s}}$ & $\mbox{Recall}_{\mbox{s}}$ & Avg & $\mbox{Accuracy}_{\mbox{s}}$ & $\mbox{Recall}_{\mbox{s}}$\\
			\hline
			\multicolumn{10}{c}{Audio + Subtitle} \\
			\hline
			Qwen-Audio(a) \cite{chu2023qwen}   & $\surd$  & $\times$ & $\surd$&40.62$\pm$0.14 & 50.03$\pm$0.22 & 31.22$\pm$0.06 & 44.90$\pm$0.06 & 55.40$\pm$0.00 & 34.40$\pm$0.11 \\
			OneLLM \cite{han2023onellm}     & $\surd$  & $\times$ & $\surd$&43.65$\pm$0.01 & 46.75$\pm$0.03 & 40.55$\pm$0.02 & 47.68$\pm$0.08 & 53.21$\pm$0.20 & 42.15$\pm$0.05 \\
			SECap \cite{xu2024secap} & $\surd$   & $\times$ & $\surd$&45.49$\pm$0.03 & 52.82$\pm$0.00 & 38.17$\pm$0.06 & 45.10$\pm$0.19 & 53.14$\pm$0.29 & 37.05$\pm$0.10 \\
			SALMONN \cite{tang2023salmonn}      & $\surd$  & $\times$ & $\surd$&47.26$\pm$0.05 & 49.21$\pm$0.05 & 45.31$\pm$0.05 & 47.93$\pm$0.22 & 51.22$\pm$0.19 & 44.63$\pm$0.25 \\
			\hline
			\multicolumn{10}{c}{Video + Subtitle} \\
			\hline
			Otter  \cite{li2023otter}   & $\surd$  & $\surd$ & $\times$&46.06$\pm$0.12 & 52.82$\pm$0.13 & 39.30$\pm$0.11 & 48.40$\pm$0.11 & 54.47$\pm$0.20 & 42.34$\pm$0.02 \\
			VideoChat  \cite{li2023videochat} & $\surd$  & $\surd$ & $\times$&45.54$\pm$0.00 & 43.25$\pm$0.15 & 47.82$\pm$0.15 & 45.79$\pm$0.11 & 47.78$\pm$0.00 & 43.80$\pm$0.22 \\
			{Video-LLaMA} \cite{zhang2023video}  & $\surd$ & $\surd$ & $\times$&{45.68}$\pm$0.13 & 45.31$\pm$0.11 & 46.05$\pm$0.15 & 47.45$\pm$0.07 & 48.42$\pm$0.09 & 46.48$\pm$0.05 \\
			Video-LLaVA \cite{lin2023video}   & $\surd$  & $\surd$ & $\times$&48.20$\pm$0.10 & 49.37$\pm$0.03 & 47.04$\pm$0.24 & 50.63$\pm$0.03 & 55.13$\pm$0.03 & 46.13$\pm$0.03 \\
			VideoChat2 \cite{li2024mvbench}   & $\surd$  & $\surd$ & $\times$&51.03$\pm$0.39 & 56.08$\pm$0.51 & 45.97$\pm$0.27 & 50.31$\pm$0.05 & 57.45$\pm$0.09 & 43.16$\pm$0.00 \\
			OneLLM \cite{han2023onellm}     & $\surd$  & $\surd$  & $\times$&50.39$\pm$0.14 & 55.25$\pm$0.16 & 45.54$\pm$0.13 & 49.86$\pm$0.10 & 54.39$\pm$0.05 & 45.33$\pm$0.14 \\
			LLaMA-VID \cite{li2023llama}    & $\surd$  & $\surd$  & $\times$&51.39$\pm$0.03 & 52.96$\pm$0.07 & 49.81$\pm$0.14 & 52.12$\pm$0.00 & 56.76$\pm$0.00 & 47.49$\pm$0.01 \\
			mPLUG-Owl \cite{ye2023mplug}   & $\surd$  & $\surd$ & $\times$&53.78$\pm$0.13 & 56.08$\pm$0.19 & 51.47$\pm$0.07 & 51.72$\pm$0.12 & 57.42$\pm$0.20 & 46.03$\pm$0.03 \\
			Video-ChatGPT \cite{maaz2023video} & $\surd$ & $\surd$ & $\times$&51.88$\pm$0.07 & 55.03$\pm$0.06 & 48.73$\pm$0.09 & 54.67$\pm$0.02 & 60.97$\pm$0.13 & 48.37$\pm$0.08 \\
			Chat-UniVi \cite{jin2023chat}   & $\surd$  & $\surd$  & $\times$&53.06$\pm$0.14 & 53.53$\pm$0.09 & 52.60$\pm$0.19 & 53.41$\pm$0.01 & 58.22$\pm$0.02 & 48.61$\pm$0.00 \\
			\hline
			\multicolumn{10}{c}{Audio + Video + Subtitle} \\
			\hline
			SECap + mPLUG-Owl       & $\surd$ & $\surd$ & $\surd$&56.07$\pm$0.02 & 48.11$\pm$0.38 & 64.02$\pm$0.35 & 54.27$\pm$0.21 & 50.73$\pm$0.18 & 57.81$\pm$0.24 \\
			SALMONN + Video-ChatGPT & $\surd$ & $\surd$ & $\surd$&58.46$\pm$0.18 & 53.09$\pm$0.04 & 63.84$\pm$0.32 & 55.17$\pm$0.05 & 52.60$\pm$0.04 & 57.74$\pm$0.14 \\
			SECap + Video-ChatGPT   & $\surd$ & $\surd$ & $\surd$&57.16$\pm$0.02 & 52.13$\pm$0.00 & 62.18$\pm$0.05 & 56.84$\pm$0.11 & 57.76$\pm$0.06 & 55.91$\pm$0.16 \\
			SECap + Chat-UniVi      & $\surd$ & $\surd$ & $\surd$&58.82$\pm$0.08 & 48.22$\pm$0.20 & 69.42$\pm$0.03 & 54.74$\pm$0.03 & 51.03$\pm$0.10 & 58.44$\pm$0.05 \\
			SALMONN + mPLUG-Owl     & $\surd$ & $\surd$ & $\surd$&58.44$\pm$0.00 & 50.91$\pm$0.08 & 65.98$\pm$0.08 & 55.27$\pm$0.18 & 51.22$\pm$0.16 & 59.33$\pm$0.19 \\
			SALMONN + Chat-UniVi    & $\surd$ & $\surd$ & $\surd$&58.69$\pm$0.04 & 50.59$\pm$0.01 & 66.79$\pm$0.09 & 57.85$\pm$0.05 & 52.51$\pm$0.05 & 63.20$\pm$0.04 \\
			\hline
			EMER(Multi) & $\surd$  & $\surd$  & $\surd$&80.23$\pm$0.25 & 79.81$\pm$0.44 & 80.65$\pm$0.06 & 84.68$\pm$0.02 & 87.02$\pm$0.09 & 82.34$\pm$0.06 \\
			\hline
		\end{tabular}
	}
\end{table}

\begin{table}[h]
	\centering
	\renewcommand\tabcolsep{2pt}
	\caption{Main results on the test set of EMER-Fine (66 samples).}
	\label{Table8}
	\scalebox{0.8}{
		\begin{tabular}{lccc|>{\columncolor{lightgray}}ccc|>{\columncolor{lightgray}}ccc}
			\hline
			\multirow{2}{*}{Model} & \multirow{2}{*}{L} & \multirow{2}{*}{V} & \multirow{2}{*}{A} & \multicolumn{3}{c|}{English} & \multicolumn{3}{c}{Chinese} \\
			&&&&Avg & $\mbox{Accuracy}_{\mbox{s}}$ & $\mbox{Recall}_{\mbox{s}}$ & Avg & $\mbox{Accuracy}_{\mbox{s}}$ & $\mbox{Recall}_{\mbox{s}}$\\
			\hline
			\multicolumn{10}{c}{Audio + Subtitle} \\
			\hline
			
			Qwen-Audio \cite{chu2023qwen}   & $\surd$  & $\times$ & $\surd$&38.66$\pm$0.13 & 46.97$\pm$0.00 & 30.35$\pm$0.25 & 38.03$\pm$0.00 & 46.97$\pm$0.00 & 29.09$\pm$0.00 \\
			OneLLM \cite{han2023onellm}     & $\surd$  & $\times$ & $\surd$&40.56$\pm$0.32 & 42.55$\pm$0.38 & 38.56$\pm$0.25 & 43.09$\pm$0.35 & 47.47$\pm$0.51 & 38.70$\pm$0.19 \\
			SECap \cite{xu2024secap} & $\surd$   & $\times$ & $\surd$&52.78$\pm$0.63 & 61.36$\pm$0.76 & 44.19$\pm$0.51 & 55.05$\pm$0.00 & 65.15$\pm$0.00 & 44.95$\pm$0.00 \\
			SALMONN \cite{tang2023salmonn}      & $\surd$  & $\times$ & $\surd$&51.28$\pm$0.00 & 54.17$\pm$0.00 & 48.38$\pm$0.00 & 50.93$\pm$0.76 & 56.31$\pm$0.76 & 45.56$\pm$0.76 \\
			
			\hline
			\multicolumn{10}{c}{Video + Subtitle} \\
			\hline
			
			Otter  \cite{li2023otter}   & $\surd$  & $\surd$ & $\times$&37.72$\pm$0.00 & 42.22$\pm$0.00 & 33.22$\pm$0.00 & 41.05$\pm$0.24 & 45.45$\pm$0.00 & 36.64$\pm$0.48 \\
			VideoChat  \cite{li2023videochat} & $\surd$  & $\surd$ & $\times$&46.34$\pm$0.45 & 41.49$\pm$0.77 & 51.19$\pm$0.13 & 44.98$\pm$0.62 & 44.90$\pm$0.61 & 45.06$\pm$0.64 \\
			{Video-LLaMA} \cite{zhang2023video}  & $\surd$ & $\surd$ & $\times$&{40.97}$\pm$0.15 & 39.44$\pm$0.18 & 42.50$\pm$0.13 & 46.55$\pm$0.13 & 46.25$\pm$0.00 & 46.84$\pm$0.25 \\
			Video-LLaVA \cite{lin2023video}   & $\surd$  & $\surd$ & $\times$&42.75$\pm$0.35 & 45.38$\pm$0.20 & 40.13$\pm$0.51 & 45.44$\pm$0.38 & 49.24$\pm$0.00 & 41.64$\pm$0.76 \\
			VideoChat2 \cite{li2024mvbench}   & $\surd$  & $\surd$ & $\times$&43.83$\pm$0.16 & 49.24$\pm$0.00 & 38.42$\pm$0.32 & 48.29$\pm$0.47 & 55.81$\pm$0.76 & 40.77$\pm$0.19 \\
			OneLLM \cite{han2023onellm}     & $\surd$  & $\surd$  & $\times$&53.40$\pm$0.19 & 58.65$\pm$0.19 & 48.14$\pm$0.19 & 59.84$\pm$0.00 & 64.65$\pm$0.00 & 55.03$\pm$0.00 \\
			LLaMA-VID \cite{li2023llama}    & $\surd$  & $\surd$  & $\times$&50.90$\pm$0.60 & 51.69$\pm$0.63 & 50.11$\pm$0.57 & 53.78$\pm$0.06 & 59.47$\pm$0.00 & 48.08$\pm$0.13 \\
			mPLUG-Owl \cite{ye2023mplug}   & $\surd$  & $\surd$ & $\times$&48.84$\pm$0.13 & 48.33$\pm$0.13 & 49.34$\pm$0.38 & 50.25$\pm$0.63 & 52.27$\pm$0.25 & 48.23$\pm$1.01 \\
			Video-ChatGPT \cite{maaz2023video} & $\surd$ & $\surd$ & $\times$&46.12$\pm$0.00 & 50.00$\pm$0.00 & 42.25$\pm$0.00 & 58.02$\pm$0.00 & 61.87$\pm$0.00 & 54.17$\pm$0.00 \\
			Chat-UniVi \cite{jin2023chat}   & $\surd$  & $\surd$  & $\times$&53.20$\pm$0.54 & 54.29$\pm$0.38 & 52.11$\pm$0.69 & 57.37$\pm$0.06 & 59.85$\pm$0.00 & 54.90$\pm$0.13 \\
			\hline
			\multicolumn{10}{c}{Audio + Video + Subtitle} \\
			\hline
			SECap + mPLUG-Owl       & $\surd$ & $\surd$ & $\surd$&64.42$\pm$0.32 & 57.95$\pm$0.38 & 70.90$\pm$0.26 & 59.00$\pm$0.25 & 55.30$\pm$0.63 & 62.70$\pm$0.13 \\
			SALMONN + Video-ChatGPT & $\surd$ & $\surd$ & $\surd$&59.71$\pm$0.47 & 53.48$\pm$0.69 & 65.93$\pm$0.25 & 54.82$\pm$0.63 & 56.82$\pm$0.88 & 52.83$\pm$0.38 \\
			SECap + Video-ChatGPT   & $\surd$ & $\surd$ & $\surd$&58.43$\pm$0.35 & 51.60$\pm$0.19 & 65.26$\pm$0.51 & 55.11$\pm$0.35 & 51.45$\pm$0.32 & 58.76$\pm$0.38 \\
			SECap + Chat-UniVi      & $\surd$ & $\surd$ & $\surd$&60.38$\pm$0.07 & 51.39$\pm$0.66 & 69.37$\pm$0.53 & 63.71$\pm$0.83 & 57.98$\pm$0.74 & 69.43$\pm$0.91 \\
			SALMONN + mPLUG-Owl     & $\surd$ & $\surd$ & $\surd$&65.15$\pm$0.26 & 55.28$\pm$0.26 & 75.03$\pm$0.26 & 58.57$\pm$0.35 & 53.78$\pm$0.32 & 63.36$\pm$0.38 \\
			SALMONN + Chat-UniVi    & $\surd$ & $\surd$ & $\surd$ &{62.64}$\pm$0.22 & 55.84$\pm$0.06 & 69.44$\pm$0.38 & 56.25$\pm$0.11 & 48.17$\pm$0.09 & 64.33$\pm$0.13 \\
			\hline
			EMER (Multi) & $\surd$  & $\surd$  & $\surd$&79.31$\pm$0.19 & 80.91$\pm$0.13 & 77.70$\pm$0.25 & 87.29$\pm$0.19 & 87.37$\pm$0.38 & 87.20$\pm$0.00 \\
			\hline
		\end{tabular}
	}
\end{table}

\end{document}